\documentclass[11pt]{article}
\usepackage{graphicx}
\usepackage{amsmath}
\usepackage{amssymb}
\usepackage{amsxtra}
\usepackage{bm}

\title{The gauge coupling unification in Grand Unified Theories based on the group $E_8$}
\author{K.V. Stepanyantz\\ Moscow State University,\\ Faculty of Physics, Department of Theoretical Physics,\\ 119991, Moscow, Russia,\\ stepan@m9com.ru}

\begin{document}

\newcommand*\xbar[1]{%
  \hbox{%
    \vbox{%
      \hrule height 0.5pt % The actual bar
      \kern0.5ex%         % Distance between bar and symbol
      \hbox{%
        \kern-0.1em%      % Shortening on the left side
        \ensuremath{#1}%
        \kern-0.1em%      % Shortening on the right side
      }%
    }%
  }%
}

\maketitle

\begin{abstract}
We consider a theory with the gauge group $E_8$ assuming that the gauge symmetry breaking pattern is $E_8 \to E_7 \times U_1 \to E_6 \times U_1 \to SO_{10} \times U_1 \to SU_5 \times U_1 \to
SU_3 \times SU_2 \times U_1$ and vacuum expectation values are acquired only by components of the representations 248. It is demonstrated that in this case there are several options for the relations between the gauge couplings of the resulting theory, but only one of them gives $\alpha_3 = \alpha_2$ and $\sin^2\theta_W = 3/8$. Also, it is the only option for which the resulting theory can include all MSSM superfields.
\end{abstract}

\allowdisplaybreaks

\section{Introduction}
\hspace*{\parindent}

Supersymmetric extensions of the Standard Model are one of the best candidates for describing the physics beyond it \cite{Mohapatra:1986uf}. The Minimal Supersymmetric Standard Model (MSSM) is the simplest of these extensions. It is a gauge theory with the group $SU_3\times SU_2\times U_1$ and softly broken supersymmetry. Consequently, MSSM contains 3 gauge coupling constants $e_3$, $e_2$, and $e_1$. (The number of gauge coupling constants is equal to the number of factors in the gauge group.) Quarks, leptons, and Higgs fields are components of the chiral matter superfields which are collected in Table \ref{Table_MSSM_Quantum_Numbers}. In this table we also present their quantum numbers with respect to $SU_3$, $SU_2$, and $U_1$ subgroups (the representations for $SU_3$ and $SU_2$ and the hypercharge for $U_1$). Also we indicate that there are 3 generations of the chiral superfields which include quarks and leptons.

\begin{table}
\hspace*{1mm}
\begin{tabular}{|c|c|c|c||c|c|c|c|}
\hline
Superfield & $SU_3$ & $SU_2$ & $U_1$ ($Y$) & Superfield & $SU_3$ & $SU_2$ & $U_1\vphantom{\Big(}$ ($Y$)\\
\hline
$3\times Q$ & \,\xbar{3}\, & 2 & $-1/6$ & $3\times N$ & 1 & 1 & $0\vphantom{\Big(}$  \\
\hline
$3\times U$ & 3 & 1 & $2/3$ & $3\times E$ & 1 & 1 & $-1\vphantom{\Big(}$\\
\hline
$3\times D$ & 3 & 1 & $-1/3$ & $H_d$ & 1 & 2 & $1/2\vphantom{\Big(}$\\
\hline
$3\times L$ & 1 & 2 & $1/2$ & $H_u$ & 1 & 2 & $-1/2\vphantom{\Big(}$\\
\hline
\end{tabular}
\caption{Quantum numbers of the various MSSM chiral matter superfields.} \label{Table_MSSM_Quantum_Numbers}
\end{table}

\noindent
Note that for the superfields which include left quarks and leptons we use the brief notations

\begin{equation}
Q = \left(
\begin{array}{c}
\widetilde U\\ \widetilde D
\end{array}
\right);\qquad L = \left(
\begin{array}{c}
\widetilde N\\ \widetilde E
\end{array}
\right).
\end{equation}

It is important that the quantum numbers of various MSSM superfields are not accidental. In particular, they satisfy the anomaly cancellation conditions, see ,e.g., \cite{Minahan:1989vd,Bilal:2008qx},

\begin{equation}
\mbox{tr}\Big(T^A\{T^B,T^C\}\Big) = 0,
\end{equation}

\noindent
where $T^A$ are the generators of the representation in which the chiral matter superfields lie. This equation is needed for the renormalizability of the theory. In MSSM this condition should be verified for all (10) possible ways of placing the gauge fileds on the external lines in the triangle diagram. The nontrivial relations needed for anomaly cancellation appear for

1. two $SU_3$ gauge fields and 1 $U_1$ gauge field;

2. two $SU_2$ gauge fields and 1 $U_1$ gauge field;

3. three $U_1$ gauge fields.

\noindent
The corresponding relations needed for the anomaly cancellation are written in the form

\begin{eqnarray}
&& Y_U + Y_D + 2\, Y_{Q} = 0;\qquad 3\, Y_{Q} + Y_{L} = 0;\qquad\nonumber\\
&& 3\,Y_U^3 + 3\,Y_D^3 + Y_E^3 + 6\,Y_{Q}^3 + 2\,Y_{L}^3 = 0.\vphantom{\Big(}
\end{eqnarray}

\noindent
Therefore, there is a question if the quantum numbers of MSSM superfields are accidental, and how they appear.

The origin of the quantum numbers of various (super)fields can presumably be explained with the help of the Grand Unification idea. Similarly to the nonsupersymmetric case first considered in \cite{Georgi:1974sy}, the MSSM superfields of a single generation (including the right neutrino) can be accommodated in 3 irreducible representations of the group $SU_5$

\begin{equation}
1 + 5 + \,\xbar{10}\,
\end{equation}

\noindent
in such a way that

\begin{equation}
1 \sim N;\quad
5_i \sim \left(
\begin{array}{c}
D_1\\ D_2\\ D_3\\ \widetilde E\\ - \widetilde N
\end{array}
\right);
\quad
\,\xbar{10}\,{}^{ij} \sim \left(
\begin{array}{ccccc}
0 & U_3 & - U_2 & \widetilde U^1 & \widetilde D^1\\
- U_3 & 0 & U_1 & \widetilde U^2 & \widetilde D^2\\
U_2 & -U_1 & 0 & \widetilde U^3 & \widetilde D^3\\
- \widetilde U^1 & -\widetilde U^2 & -\widetilde U_3 & 0 & E\\
- \widetilde D^1 & -\widetilde D^2 & -\widetilde D^3 & -E & 0
\end{array}
\right).
\end{equation}

\noindent
In this case after the symmetry breaking $SU_5\to SU_3\times SU_2\times U_1$ all fields of the low-energy theory will have correct quantum numbers.

The $SU_5$ symmetry can be broken down to the subgroup $SU_3\times SU_2\times U_1$ with the elements

\begin{equation}
\omega_5 = \left(
\begin{array}{cc}
\omega_3 e^{-i\beta_Y/3} & 0 \\
0 & \omega_2^* e^{i\beta_Y/2}
\end{array}
\right) \in SU_3\times SU_2 \times U_1 \subset SU_5.
\end{equation}

\noindent
by a vacuum expectation value (vev) of the Higgs field in the adjoint representation $24$. Then, from the $SU_5$ tensor transformations

\begin{equation}
N \to N;\qquad 5_i \to (\omega_5)_i{}^j 5_j;\qquad \,\xbar{10}\,{}^{ij} \to (\omega_5^*)^i{}_k (\omega_5^*)^j{}_l \,\xbar{10}\,{}^{kl}
\end{equation}

\noindent
one obtains that with respect to the subgroup $SU_3\times SU_2\times U_1$ all chiral superfields have the same quantum numbers as the MSSM superfields. The further symmetry breaking $SU_3\times SU_2\times U_1 \to SU_3\times U_{1}^{em}$ is usually made by vacuum expectation values of the Higgs superfields in the representations $5$ and $\,\xbar{5}\,$. However, in this case the doublet-triplet splitting requires fine tuning \cite{Dimopoulos:1981zb,Sakai:1981gr}.

The anomaly cancellation in this model occurs due to the relation

\begin{equation}
\mbox{tr}\big(T^A\{T^B,T^C\}\big)\Big|_5 + \mbox{tr}\big(T^A\{T^B,T^C\}\big)\Big|_{\overline{10}} = 0.
\end{equation}

Because the group $SU_5$ is simple, there is the only gauge coupling constant $e_5$ in the $SU_5$ Grand Unified Theory (GUT). This implies that in the low-energy $SU_3\times SU_3\times U_1$ theory 3 coupling constants should be related to each other. This relation is written as

\begin{equation}\label{Coupling_Relations}
\alpha_2=\alpha_3; \qquad \sin^2\theta_W=3/8,
\end{equation}

\noindent
where $\mbox{tg}\,\theta_W \equiv e_1/e_2$. If we introduce the notation ${\displaystyle \alpha_1\equiv \frac{5}{3}\cdot \frac{e_1^2}{4\pi}}$, then the gauge coupling unification condition takes the simplest form $\alpha_1=\alpha_2=\alpha_3 = \alpha_5$. This condition is in a good agreement with the well-known renormalization group behaviour of the running gauge couplings in MSSM \cite{Ellis:1990wk,Amaldi:1991cn,Langacker:1991an}.

The field content of the $SU_5$ GUT indicates on a possibility of the existence of a theory with a wider $SO_{10}$ symmetry \cite{Fritzsch:1974nn,Georgi:1974my} because the superfields of a single generation can arise from a single irreducible (spinor) $SO_{10}$ representation

\begin{equation}\label{16_Decomposion}
\,\xbar{16}\,\Big|_{SO_{10}}=1(5)+5(-3)+\,\xbar{10}\,(1)\Big|_{SU_5\times U_1}.
\end{equation}

\noindent
However, the symmetry breaking pattern

\begin{equation}
SO_{10} \to SU_5 \to SU_3\times SU_2\times U_1
\end{equation}

\noindent
has some drawbacks. In particular, for the symmetry breaking one needs (super)fields in sufficiently large representations (no less than $45$ of $SO_{10}$ and $24$ of $SU_5$). Moreover, the simplest (supersymmetric) $SU_5$ model is excluded by the modern experimental limits on the proton lifetime \cite{Workman:2022ynf}.  A more convenient symmetry breaking pattern is

\begin{equation}
SO_{10}\to SU_5\times U_1 \to SU_3\times SU_2\times U_1.
\end{equation}

\noindent
It corresponds to the flipped $SU_5$ model \cite{Barr:1981qv,Antoniadis:1987dx,Campbell:1987eb,Ellis:1988tx}. In this case the chiral matter superfields containing quarks and leptons are situated in the representations $3\times \Big(\overline{10} (1) + 5 (-3) + 1 (5)\Big)$ in a different way,

\begin{equation}
1\sim E;\quad 5_i \sim \left(\begin{array}{c} U_1\\ U_2\\ U_3\\ \widetilde E\\ -\widetilde N
\end{array}\right);\quad \overline{10}^{ij} \sim \left(
\begin{array}{ccccc}
0 & D_3 & -D_2 & \widetilde U^1 & \widetilde D^1\\
-D_3 & 0 & D_1 & \widetilde U^2 & \widetilde D^2\\
D_2 & -D_1 & 0 & \widetilde U^3 & \widetilde D^3\\
-\widetilde U^1 & -\widetilde U^2 & - \widetilde U^3 & 0 & N\\
-\widetilde D^1 & -\widetilde D^2 & -\widetilde D^3 & -N & 0
\end{array}
\right),
\end{equation}

\noindent
so that the superfields corresponding to the right up and down quarks and leptons are swapped. (That is why this model is called ``flipped''.)

The $SU_5\times U_1$ symmetry is broken down to $SU_3\times SU_2 \times U_1^{(Y)}$ by vevs of Higgses in the representations $10 (-1)$ and $\,\xbar{10}\,(1)$, and the group $U_1^{(Y)}$ appears as a superposition of the $SU(5)$ transformations with

\begin{equation}
\omega_5 = \exp\Big\{\,\frac{i\alpha_Y}{30}\left(
\begin{array}{cc}
2\cdot 1_3 & 0 \\
0 & -3\cdot 1_2
\end{array}
\right)\Big\}
\end{equation}

\noindent
and the $U_1$ transformations with $\omega_1 = \exp(-iQ\alpha_Y/5)$, where $Q$ is the $U_1$ charge normalized by Eq. (\ref{16_Decomposion}). This model

1. allows to naturally split Higgs douplet and triplet \cite{Antoniadis:1987dx,Masiero:1982fe,Grinstein:1982um,Hisano:1994fn};

2. does not require higher representations for the breaking of the $SU_5$ symmetry;

3. satisfies present limits on the proton lifetime \cite{Ellis:2020qad} (see also \cite{Mehmood:2020irm,Haba:2021rzs,Ellis:2021vpp}).

\smallskip

The flipped $SU_5$ model has 2 coupling constants $e_5$ and $e_1$. However, if it is considered as a remnant of the $SO_{10}$ theory, then they are related to each other as

\begin{equation}
e_5 = \frac{e_{10}}{\sqrt{2}};\qquad e_1 = \frac{e_5}{2\sqrt{10}} = \frac{e_{10}}{4\sqrt{5}}.
\end{equation}

\noindent
Then for the residual $SU_3\times SU_2\times U_1$ theory we obtain the standard relations (\ref{Coupling_Relations}).

\begin{figure}[h]
\includegraphics[scale=0.8]{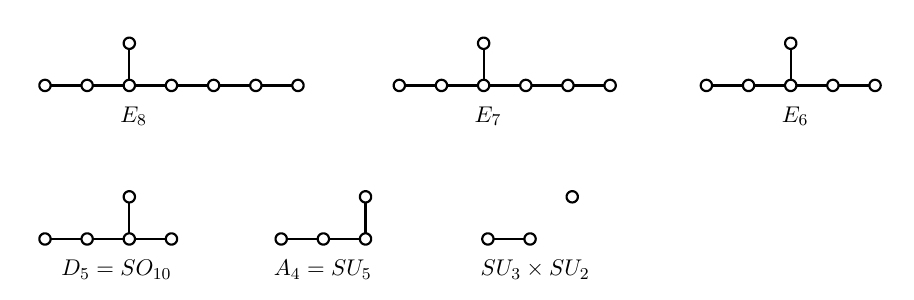}
\caption{Dynkin diagrams for the Lie algebras of the $E$-series with some classical algebras included}\label{Figure_Dynkin}
\end{figure}

Also for constructing various GUTs it is possible to consider larger groups, for example, the exceptional group $E_6$ \cite{Gursey:1975ki} (see \cite{King:2020ldn} for a review) or even $E_8$ \cite{Konshtein:1980km,Baaklini:1980uq,Baaklini:1980fv,Bars:1980mb,Koca:1981xd,Thomas:1985be}. Using of the exceptional groups $E_7$ and $E_8$ is complicated by the fact that they have only real representation and the corresponding theories are not chiral \cite{Slansky:1981yr}. However, it is known that the Lie algebras used for constructing various GUTs can be considered as a part of the $E$-series if we also include in it some classical Lie algebras, see Fig. \ref{Figure_Dynkin}.

Here (following \cite{Stepanyantz:2023vat}) we will investigate a possibility of constructing GUT based on the group $E_8$, which is the largest of them, assuming that the symmetry breaking pattern is

\begin{equation}\label{Symmetry_Breaking_Pattern}
E_8\to E_7\times U_1 \to E_6\times U_1 \to SO_{10}\times U_1 \to SU_5 \times U_1 \to SU_3 \times SU_2 \times U_1,
\end{equation}

\noindent
and vevs responsible for the various symmetry breakings are acquired only by certain parts of the fundamental representation of the group $E_8$ (of the dimension $248$). Namely, we will investigate the unification of the gauge couplings and study a possibility of obtaining the conditions (\ref{Coupling_Relations}).

The paper is organized as follows. In section \ref{Section_E_Algebras} we present explicit commutation relations for the exceptional Lie algebras of the $E$-series. In the next section \ref{SEction_Symmetry_Breaking} we discusss the symmetry breaking by vevs of scalar fields in various parts of the representation 248 of $E_8$. The relations between the coupling constants of the non-Abelian groups are derived in section \ref{Section_Non_Abelian_Relations}. The similar relations of the $U_1$ coupling constants for various options of the symmetry breaking are discussed in section \ref{Section_Options}. Conclusion contains a brief summary of the results.

\section{Explicit commutation relations for the exceptional Lie algebras of the $E$-series}
\hspace*{\parindent}\label{Section_E_Algebras}

For investigating the symmetry breaking pattern (\ref{Symmetry_Breaking_Pattern}) we will use the explicit construction of all Lie algebras entering it. In particular, we will need the explicit commutation relations for the exceptional algebras $E_8$, $E_7$, and $E_6$, which will be presented in this section.

\subsection{The $\Gamma$-matrices in diverse dimensions}
\hspace*{\parindent}

For constructing the commutation relations for the exceptional Lie algebras in the explicit form we will need the $\Gamma$-matrices in the space of a dimension $D$ and the Euclidean signature. By definition, they satisfy the condition

\begin{equation}
\{\Gamma_i,\Gamma_j\} = 2\delta_{ij}\cdot 1.
\end{equation}

\noindent
It is known that for an even $D$ they have the size $2^{D/2}\times 2^{D/2}$, and for an odd $D$ their size is $2^{(D-1)/2}\times 2^{(D-1)/2}$. In particular, for $D=2,3$ as the $\Gamma$-matrixes one can choose the Pauli matrices,

\begin{equation}
\Gamma_1^{(2)} = \sigma_1;\quad \Gamma_2^{(2)} = \sigma_2;\qquad\quad
\Gamma_1^{(3)} = \sigma_1;\quad \Gamma_2^{(3)} = \sigma_2;\quad \Gamma_3^{(3)} = \sigma_3.
\end{equation}

\noindent
For larger values of $D$ the $\Gamma$-matrices are constructed with the help of mathematical induction. Suppose that we have constructed them in an odd dimension $D$. Then in the next (even) dimension $D+1$ the $\Gamma$-matrices
have the size in two times larger and can be chosen in the form

\begin{equation}
\Gamma^{(D+1)}_i = \left(
\begin{array}{cc}
0 & \Gamma^{(D)}_i\\
\Gamma^{(D)}_i & 0
\end{array}
\right),\quad i=1,\ldots,D;\qquad \Gamma^{(D+1)}_{D+1} = \left(
\begin{array}{cc}
0 & -i\\
i & 0
\end{array}
\right).
\end{equation}

\noindent
In this case it is also possible to construct the matrix

\begin{equation}
\Gamma^{(D+1)}_{D+2} = \left(
\begin{array}{cc}
1 & 0 \\
0 & -1
\end{array}
\right),
\end{equation}

\noindent
which satisfies the conditions

\begin{equation}
\big\{\Gamma^{(D+1)}_{D+2}, \Gamma^{(D+1)}_{i}\big\} = 0,\quad i=1,\ldots D+1;\qquad \big(\Gamma^{(D+1)}_{D+2}\big)^2 =1.
\end{equation}

\noindent
This implies that in the next (odd) dimension $D+2$ as $\Gamma$-matrices one can take the $\Gamma$-matrices from the previous (even) dimension supplemented by the matrix $\Gamma^{(D+1)}_{D+2}$,

\begin{equation}
\Gamma_i^{(D+2)}\equiv \Gamma_i^{(D+1)},\quad \mbox{where}\quad i=1,\ldots D+2.
\end{equation}

\noindent
Thus, the induction step is completed.

The above constructed $\Gamma$-matrices are Hermitian, $(\Gamma_i)^+ = \Gamma_i$. Therefore, for odd $i$ they are symmetric, while for even $i$ they are antisymmetric. In an even dimension $D$ the charge conjugation matrix is defined as

\begin{equation}
\smash{C \equiv \Gamma_1 \Gamma_3 \ldots \Gamma_{D-1}}
\end{equation}

\noindent
and satisfies the conditions

\begin{equation}
C\,\Gamma_i\, C^{-1} = - (-1)^{D/2}\, (\Gamma_i)^T;\qquad C^{-1} = C^+ = C^T = (-1)^{D(D-2)/8}\, C.
\end{equation}

\noindent
Consequently, for the antisymmetrized products of $\Gamma$-matrixes the identities

\begin{eqnarray}
&& (\Gamma_{i_1 i_2\ldots i_k} C)^T = (-1)^{(D-2k)(D-2k-2)/8}\, \Gamma_{i_1 i_2\ldots i_k} C;\vphantom{\Big(}\nonumber\\
&& (\Gamma_{i_1 i_2\ldots i_k} \Gamma_{D+1} C)^T = (-1)^{(D-2k)(D-2k+2)/8}\, \Gamma_{i_1 i_2\ldots i_k} \Gamma_{D+1} C\qquad
\end{eqnarray}

\noindent
are valid in the case of even $D$.

\subsection{Notations}
\hspace*{\parindent}

We will denote the generators of the fundamental representation by $t_{\bm{A}}$, where $\bm{A}=1,\ldots,\text{dim}\, G$. They are normalized by the condition

\begin{equation}
\mbox{tr} (t_{\bm{A}} t_{\bm{B}}) = \frac{1}{2} g_{\bm{AB}},
\end{equation}

\noindent
where $g_{\bm{AB}}$ is a (symmetric) metric, and $g^{\bm{AB}}$ is the corresponding inverse matrix.

For an arbitrary representation $R$ the generators $T_{\bm{A}}$ satisfy the equations

\begin{equation}
\mbox{tr} (T_{\bm{A}} T_{\bm{B}}) = T(R) g_{\bm{AB}};\qquad [T_{\bm{A}}, T_{\bm{B}}] = i f_{\bm{AB}}{}^{\bm{C}} T_{\bm{C}},
\end{equation}

\noindent
where $f_{\bm{AB}}{}^{\bm{C}}$ are the structure constants. The expression $f_{\bm{ABC}}\equiv g_{\bm{CD}} f_{\bm{AB}}{}^{\bm{D}}$ is totally antisymmetric, and

\begin{equation}
(T_{Adj\,\bm{A}})^{\bm{C}}{}_{\bm{B}} = if_{\bm{AB}}{}^{\bm{C}};\qquad C_2\, g_{\bm{AB}} \equiv - f_{\bm{AC}}{}^{\bm{D}} f_{\bm{BD}}{}^{\bm{C}}.
\end{equation}

\noindent
In particular, from these equations we obtain $g^{\bm{AB}} [T_{\bm{A}}, [T_{\bm{B}}, T_{\bm{C}}]] = C_2 T_{\bm{C}}$. Also we note that for irreducible representations

\begin{equation}
C(R)_{\bm{i}}{}^{\bm{j}}\equiv g^{\bm{AB}} (T_{\bm{A}} T_{\bm{B}})_{\bm{i}}{}^{\bm{j}} = C(R)\,\delta_{\bm{i}}{}^{\bm{j}},\quad \mbox{where}\quad C(R) = T(R)\cdot \frac{\mbox{dim}\,G}{\mbox{dim}\,R}.
\end{equation}

\subsection{The group $E_8$}
\hspace*{\parindent}

According to \cite{Slansky:1981yr}, the fundamental representation of the group $E_8$ coincides with the adjoint representation and has the dimension 248. The group $E_8$ has the maximal subgroup $SO_{16}\subset E_8$, with respect to which

\begin{equation}
248\Big|_{E_8} = 120+128\Big|_{SO_{16}},
\end{equation}

\noindent
where 120 is the adjoint representation of $SO_{16}$, and 128 is its representation by Majorana-Weyl (right, for the definiteness) spinors. Therefore, the $E_8$ generators can be written as the set \cite{Green:1987sp}

\begin{equation}
t_{\bm{A}} = \big\{t_a,\, t_{ij}\big\},
\end{equation}

\noindent
where $i,j=1,\ldots,16$ and $a=1,\ldots,128$. The commutation relations of the group $E_8$ can be presented in the form

\begin{eqnarray}
&& [t_{ij},t_{kl}] = \frac{i}{\sqrt{120}}\Big(\delta_{il} t_{jk} - \delta_{jl} t_{ik} - \delta_{ik} t_{jl} + \delta_{jk} t_{il}\Big);\nonumber\\
&& [t_{ij},t_a] = -\frac{i}{\sqrt{480}} (\Gamma^{(16)}_{ij})_a{}^b t_b;\nonumber\\
&& [t_a,t_b] = - \frac{i}{2\sqrt{480}} (\Gamma^{(16)}_{ij} C^{(16)})_{ab} t_{ij}.
\end{eqnarray}

\noindent
where charge conjugation matrix in $D=16$ denoted by $C^{(16)}$ is symmetric, and the matrices $\Gamma^{(16)}_{ij} C^{(16)}$ are antisymmetric. The corresponding metric has the form

\begin{equation}
g^{\bm{AB}} \to \left(
\begin{array}{cc}
{\displaystyle \frac{1}{4}\big(\delta_{ik}\delta_{jl}-\delta_{il}\delta_{jk}\big)} & 0\\
0 & (C^{(16)})^{ab}
\end{array}
\right).
\end{equation}

\noindent
In particular, it is easy to verify the identity

\begin{equation}
g^{\bm{AB}} t_{\bm{A}} t_{\bm{B}} = \frac{1}{2} t_{ij} t_{ij} + (C^{(16)})^{ab} t_a t_b = \frac{1}{2}.
\end{equation}

\subsection{The group $E_7$}
\hspace*{\parindent}

To describe the group $E_7$, it is convenient to use its subgroup $SO_{12}\times SO_3$. Then \cite{Slansky:1981yr}

\begin{eqnarray}
&& 56\Big|_{E_7} = [12,2] + [32,1]\Big|_{SO_{12}\times SO_3};\nonumber\\
&& 133\Big|_{E_7} = [1,3] + [32',2] + [66,1]\Big|_{SO_{12}\times SO_3},
\end{eqnarray}

\noindent
where $32$ and $32'$ are right and left spinor representations of $SO_{12}$. The indices of the left $SO_{12}$ spinors are denoted by dots. Therefore, the $E_7$ generators can be written as the set

\begin{equation}
t_{\bm{A}} = \big\{t_{ij}, t_\alpha, t_{a\dot{A}} \big\},
\end{equation}

\noindent
where $a,b=1,2$;\ $\alpha,\beta=1,\ldots,3$;\ $i,j=1,\ldots,12$. Their commutation relations take the form \cite{Stepanyantz:2023vat}

\begin{eqnarray}
&&\hspace*{-4mm} [t_\alpha,t_\beta] = \frac{i}{\sqrt{12}} \varepsilon_{\alpha\beta\gamma} t_\gamma;\qquad [t_\alpha,t_{ij}] = 0;\nonumber\\
&&\hspace*{-4mm} [t_\alpha,t_{a\dot{A}}] = -\frac{1}{2\sqrt{12}} (\sigma_\alpha)_a{}^b t_{b\dot{A}};\qquad [t_{ij}, t_{a\dot{A}}] = - \frac{i}{2\sqrt{24}} (\Gamma^{(12)}_{ij})_{\dot{A}}{}^{\dot{B}} t_{a\dot{B}};\nonumber\\
&&\hspace*{-4mm} [t_{ij},t_{kl}] = \frac{i}{\sqrt{24}}\Big(\delta_{il} t_{jk} - \delta_{jl} t_{ik} - \delta_{ik} t_{jl} + \delta_{jk} t_{il}\Big);\\
&&\hspace*{-4mm} [t_{a\dot{A}}, t_{b\dot{B}}] = \frac{i}{4\sqrt{24}} (\sigma_2)_{ab} (\Gamma^{(12)}_{ij} C^{(12)})_{\dot{A}\dot{B}} t_{ij} + \frac{1}{2\sqrt{12}} (C^{(12)})_{\dot{A}\dot{B}} (\sigma_\alpha\sigma_2)_{ab} t_\alpha.\nonumber
\end{eqnarray}

\noindent
and the corresponding metric is

\begin{equation}
g^{\bm{AB}} \to \left(
\begin{array}{ccc}
{\displaystyle \frac{1}{4}\big(\delta_{ik}\delta_{jl}-\delta_{il}\delta_{jk}\big)} & 0 & 0\\
0 & \delta_{\alpha\beta} & 0\\
0 & 0 & (\sigma_2)^{ab} (C^{(12)})^{\dot{A}\dot{B}}
\end{array}
\right).
\end{equation}

\noindent
Note that the matrices $\sigma_2 = i\sigma_1\sigma_3$ and $C^{(12)}$ are antisymmetric, while the matrices $\sigma_\alpha\sigma_2$ and $\Gamma^{(12)}_{ij} C^{(12)}$ are symmetric, so that this metric is really symmetric.

In the explicit form the generators of the fundamental representation 56 are written as

\begin{eqnarray}
&& t_{ij} = \frac{i}{\sqrt{24}}
\left(
\begin{array}{cc}
(\delta_{ik}\delta_{jl} - \delta_{il}\delta_{jk})\delta_a^b & 0\\
0 & {\displaystyle \frac{1}{2}}(\Gamma^{(12)}_{ij})_{A}{}^{B}
\end{array}
\right);\nonumber\\
&& t_\alpha = \frac{1}{2\sqrt{12}} \left(
\begin{array}{cc}
\delta_{kl} (\sigma_\alpha)_a{}^b & 0\\
0 & 0
\end{array}
\right);\nonumber\\
&& t_{d\dot{D}} = \frac{i}{2\sqrt{12}} \left(
\begin{array}{cc}
0 & (\sigma_2)_{da} (\Gamma^{(12)}_k)_{\dot{D}}{}^{B}\\
(\Gamma^{(12)}_l C^{(12)})_{A\dot{D}} \delta_d^b & 0
\end{array}
\right).\qquad
\end{eqnarray}

\noindent
As a check, one can verify that

\begin{equation}
C(56) = \frac{1}{2} t_{ij} t_{ij} + t_\alpha t_\alpha + (\sigma_2)^{ab} (C^{(12)})^{\dot{A}\dot{B}} t_{a\dot{A}} t_{b\dot{B}} = \frac{19}{16} = \frac{1}{2}\cdot \frac{133}{56}.
\end{equation}

\subsection{The group $E_6$}
\hspace*{\parindent}

For describing the group $E_6$ we will use its maximal subgroup $SO_{10}\times U_1$ with respect to that

\begin{eqnarray}\label{E6_Branching_Rules}
&& 27\Big|_{E_6} = 1(4) + 10(-2)+16(1)\Big|_{SO_{10}\times U_1};\nonumber\\
&& \,\xbar{27}\,\Big|_{E_6} = 1(-4) + 10(2)+\,\xbar{16}\,(-1)\Big|_{SO_{10}\times U_1};\nonumber\\
&& 78\Big|_{E_6} = 1(0) + 16(-3) + \,\xbar{16}\,(3) + 45(0)\Big|_{SO_{10}\times U_1},
\end{eqnarray}

\noindent
where $16$ and $\xbar{16}$ are the right and left spinor representations of $SO_{10}$. However, now we will use a single spinor index $a=1,\ldots,32$, so that the $E_6$ generators can presented as the set

\begin{equation}
t_{\bm{A}} = \big\{t_{ij}, t_a, t\big\}.
\end{equation}

\noindent
In this case their commutation relations are written in the form \cite{Stepanyantz:2023vat}

\begin{eqnarray}
&& [t_{ij},t_{kl}] = \frac{i}{\sqrt{12}}\Big(\delta_{il} t_{jk} - \delta_{jl} t_{ik} - \delta_{ik} t_{jl} + \delta_{jk} t_{il}\Big);\nonumber\\
&& [t_{ij},t] = 0;\qquad [t,t_a] = \frac{1}{4} (\Gamma^{(10)}_{11})_a{}^b t_b;\qquad [t_{ij},t_a] = - \frac{i}{2\sqrt{12}} (\Gamma^{(10)}_{ij})_a{}^b t_b;\nonumber\\
&& [t_a,t_b] = - \frac{i}{4\sqrt{12}} (\Gamma^{(10)}_{ij}C^{(10)})_{ab} t_{ij} + \frac{1}{4} (\Gamma^{(10)}_{11} C^{(10)})_{ab} t,
\end{eqnarray}

\noindent
and the corresponding metric is

\begin{equation}
g^{\bm{AB}} \to \left(
\begin{array}{ccc}
{\displaystyle \frac{1}{4}\big(\delta_{ik}\delta_{jl}-\delta_{il}\delta_{jk}\big)} & 0 & 0\\
0 & (C^{(10)})^{ab} & 0\\
0 & 0 & 1
\end{array}
\right).
\end{equation}

\noindent
Note that in $D=10$ the matrix $C^{(10)}$ is symmetric and coincides with its inverse, while the matrices $\Gamma^{(10)}_{ij} C^{(10)}$ and $\Gamma^{(10)}_{11} C^{(10)}$ are antisymmetric.

In the explicit form the generators of the fundamental representation $27$ are written as

\footnotesize

\begin{eqnarray}
&& t_{ij} = \frac{i}{\sqrt{12}} \left(
\begin{array}{ccc}
0 & 0 & 0\\
0 & \delta_{ik}\delta_{jl} - \delta_{il}\delta_{jk} & 0\vphantom{\Big(_a}\\
0 & 0 & {\displaystyle \frac{1}{4}} \left[\Gamma^{(10)}_{ij}(1+\Gamma^{(10)}_{11})\right]_a{}^b
\end{array}
\right);\nonumber\\
&& t = \frac{1}{12}
\left(
\begin{array}{ccc}
4 & 0 & 0\vphantom{\Big(_a}\\
0 & -2\delta_{kl} & 0\\
0 & 0 & {\displaystyle \frac{1}{2}}(1+\Gamma^{(10)}_{11})_a{}^b
\end{array}
\right);\qquad \\
&& \hspace*{-8mm}
t_d = \frac{1}{\sqrt{96}}\left(
\begin{array}{ccc}
0 & 0 & \sqrt{2} \big(1+\Gamma^{(10)}_{11}\big)_d{}^b\vphantom{\Big(_a}\\
0 & 0 & \left[\Gamma^{(10)}_k(1+\Gamma^{(10)}_{11})\right]_d{}^b\\
\sqrt{2}\left[(1+\Gamma^{(10)}_{11}) C^{(10)}\right]_{ad} & \left[(1+\Gamma^{(10)}_{11}) \Gamma^{(10)}_l C^{(10)}\right]_{ad} & 0
\end{array}
\right)\nonumber
\end{eqnarray}

\normalsize

\noindent
Again, as a check, one can verify that

\begin{equation}
C(27) = \frac{1}{2} t_{ij} t_{ij} + (C^{(10)})^{ab} t_a t_b + t^2 = \frac{13}{9} = \frac{1}{2}\cdot \frac{78}{27}.
\end{equation}

\section{The symmetry breaking}
\hspace*{\parindent}\label{SEction_Symmetry_Breaking}

In this section we will analyze the representations which can be used for realizing the symmetry breaking pattern (\ref{Symmetry_Breaking_Pattern}) and find some relations between the coupling constants which appear at various stages of this symmetry breaking. 

\subsection{The symmetry breaking $E_8\to E_7\times U_1$}
\hspace*{\parindent}

Let us investigate if it is possible to break the $E_8$ symmetry by vev of the representation $248$. For this purpose we consider the embedding

\begin{equation}
E_8 \supset SO_{16} \supset  \underbrace{SO_{12}\times SO_3}_{\subset E_7}\times SO_{3},
\end{equation}

\noindent
for which

\begin{eqnarray}
&&\hspace*{-7mm} 248\Big|_{E_8} = 120 + 128\Big|_{SO_{16}}  = \smash{\underbrace{[1,1,3]}_{[1,3]\vphantom{\Big|_{E_7\times SO_3}}} + \underbrace{[1,3,1] + [66,1,1] + [32',2,1]}_{+ [133,1]\vphantom{\Big|_{E_7\times SO_3}}}}\vphantom{\int\limits_p}\nonumber\\
&&\hspace*{-7mm} + \underbrace{[12,2,2] + [32,1,2]}_{+ [56,2]\Big|_{E_7\times SO_3}}\Big|_{SO_{12}\times SO_3\times SO_3}.\qquad
\end{eqnarray}

\noindent
Let us assume that a scalar field in the representation $248$ is responsible for the symmetry breaking and present it in the form

\begin{equation}
\varphi = \varphi_{\bm{A}}\, g^{\bm{AB}} t_{\bm{B}} = \frac{1}{2} \varphi_{ij} t_{ij} + \varphi_a (C^{(16)})^{ab} t_b
\end{equation}

\noindent
supposing that

\begin{equation}\label{E8_Vev}
\big(\varphi_{13,14}\big)_0 = \big(\varphi_{15,16}\big)_0 = v_8.
\end{equation}

Next, we construct the corresponding little group under which, by definition, a vacuum expectation remains invariant. This implies that it is necessary to find all $E_8$ generators which commute with $\varphi_0 = v_8 \big(t_{13,14} + t_{15,16}\big)$. Evidently, this vev commutes with all $t_{ij}$ with $i,j=1,\ldots,12$, which form the subgroup $SO_{12}$. Also the little group includes the generators

\begin{eqnarray}
&& \widetilde t_1 \equiv \frac{1}{\sqrt{2}} \big(t_{13,16} - t_{14,15}\big); \qquad \widetilde t_2 \equiv \frac{1}{\sqrt{2}} \big(-t_{13,15} - t_{14,16}\big);\qquad\nonumber\\
&& \widetilde t_3 \equiv \frac{1}{\sqrt{2}} \big(t_{13,14} - t_{15,16}\big); \qquad \widetilde t_3' \equiv \frac{1}{\sqrt{2}}\big(-t_{13,14} - t_{15,16}\big).\qquad\nonumber
\end{eqnarray}

\noindent
They form the subgroup $SO_3\times U_1$ of the little group,

\begin{equation}
[\widetilde t_3', \widetilde t_\alpha] = 0;\qquad [\widetilde t_\alpha,\widetilde t_\beta] = \frac{i}{\sqrt{60}} \varepsilon_{\alpha\beta\gamma} \widetilde t_\gamma.
\end{equation}

\noindent
However, the little group is wider than $SO_{12}\times SO_3\times U_1$ because some generators $t_a$ also commute with the vacuum expectation value. Really,

\begin{equation}
[\varphi_0, t_a] = v_8\, \big[t_{13,14}+t_{15,16},t_a\big] = -\frac{i v_8}{2\sqrt{120}} \big(\Gamma^{(16)}_{13,14} + \Gamma^{(16)}_{15,16}\big)_a{}^b\, t_b.
\end{equation}

\noindent
where

\begin{equation}
-\frac{i}{2}\Big(\Gamma^{(16)}_{13,14} + \Gamma^{(16)}_{15,16}\Big) = \left(
\begin{array}{cc}
1 & 0 \\
0 & -1
\end{array}
\right) \cdot \frac{1}{2}\Big(1+\Gamma^{(12)}_{13}\Big).
\end{equation}

\noindent
Therefore, the generators $t_a$ which belong to the little group form two left 32 component $SO_{12}$ spinors, which transform under the spinor representation 2 of the group $SO_3$. Thus, we obtain that the little group is $E_7\times U_1$ because

\begin{equation}
133\Big|_{E_7} = [1,3] + [32',2] + [66,1]\Big|_{SO_{12}\times SO_3}.
\end{equation}

\noindent
Therefore, by the vev (\ref{E8_Vev}) the symmetry is broken as

\begin{equation}
E_8\to E_7\times U_1.
\end{equation}

Next, let us relate 2 coupling constants of the resulting theory with the original coupling constant $e_8$. Comparing the commutation relations of the generators $t_{ij}$ for the groups $E_8$ and $E_7$ we see that

\begin{equation}
t_{ij}\Big|_{E_8} = \frac{1}{\sqrt{5}} t_{ij}\Big|_{E_7}.
\end{equation}

\noindent
Because $A_\mu = i e A_\mu^{\bm{A}} t_{\bm{A}}$ and the generators $t_{ij}$ are normalized by the same condition

\begin{equation}
\mbox{tr}(t_{ij} t_{kl}) = \frac{1}{2}\big(\delta_{ik}\delta_{jl} - \delta_{il}\delta_{jk}\big),
\end{equation}

\noindent
the coupling constants for the groups $E_7$ and $E_8$ are related as

\begin{equation}
e_7 = \frac{e_8}{\sqrt{5}}.
\end{equation}

The coupling constant $e^{(7)}_1$ corresponding to the subgroup $U_1$ depends on the normalization of the $U_1$ charge. Let us choose the $SO_3$ generators in the subgroup $E_7\times SO_3\subset E_8$ in such a way that

\begin{equation}
[t_\alpha',t_\beta'] = 2i\varepsilon_{\alpha\beta\gamma} t_\gamma'.
\end{equation}

\noindent
and take $t_3'$ as a generator of the $U_1$ component of the little group. For this normalization condition

\begin{equation}
248\Big|_{E_8} = 1(0) + 1(2) + 1(-2) + 133(0) + 56(1) + 56(-1)\Big|_{E_7\times U_1}.
\end{equation}

\noindent
From the other side, the generators of the $SO_3$ subgroup in $E_7\times SO_3\subset E_8$ normalized in the same way as all $E_8$ generators satisfy the commutation relation

\begin{equation}
[\widetilde t'_\alpha,\widetilde t'_\beta] = \frac{i}{\sqrt{60}} \varepsilon_{\alpha\beta\gamma} \widetilde t'_\gamma.
\end{equation}

\noindent
Comparing it with the commutation relation for $t'_\alpha$ we see that

\begin{equation}
\widetilde t_3' = \frac{1}{4\sqrt{15}} t_3'.
\end{equation}

\noindent
This implies that the corresponding couplings are related by the equation

\begin{equation}\label{E17_Coupling}
e_1^{(7)} = \frac{e_8}{4\sqrt{15}} = \frac{e_7}{4\sqrt{3}}.
\end{equation}

\subsection{The symmetry breaking $E_7\times U_1 \to E_6\times U_1$}
\hspace*{\parindent}

The group $E_7$ contains the maximal subgroup $E_6\times U_1$, with respect to which

\begin{eqnarray}\label{E7_Branching_Rules}
&& 56\Big|_{E_7} = 27(1) + \xbar{27}\,(-1) + 1(3) + 1(-3)\Big|_{E_6\times U_1};\nonumber\\
&& 133\Big|_{E_7} = 1(0) + 27(-2) + \xbar{27}\,(2) + 78(0)\Big|_{E_6\times U_1}.
\end{eqnarray}

\noindent
In particular, the representation $56$ contains two $E_6$ singlets with nontrivial $U_1$ charges. If one of them acquires a vacuum expectation value, then the little group will contain the factor $E_6$. Let the vacuum expectation value $v_7$ is acquired by the representation $56(1)$ of the group $E_7\times U_1$, and the corresponding scalar field lies in the representation $1(3)$ of the group $E_6\times U_1\subset E_7$. Under the $U_1\times U_1$ transformations in

\begin{equation}
E_7\times \underbrace{U_1}_{\beta_1^{(7)}} \supset (E_6\times \underbrace{U_1}_{\beta_2^{(7)}})\times \underbrace{U_1}_{\beta_1^{(7)}}.
\end{equation}

\noindent
the vacuum expectation value changes as $v_7 \to \exp\Big(i\beta^{(7)}_1 + 3i\beta^{(7)}_2\Big)\, v_7$. Therefore, it is invariant under the transformations with $\beta^{(7)}_1 + 3\beta^{(7)}_2 = 0$. Evidently, they constitute the group $U_1 \subset U_1\times U_1$. Therefore, the little group in this case is $E_6\times U_1$.

Next, we compare the coupling constants in the original $E_7\times U_1$ theory and in its $E_6\times U_1$ remnant. As earlier, comparing the commutation relations for the generators $t_{ij}$ we find the relation between the couplings for the groups $E_7$ and $E_6$,

\begin{equation}
t_{ij}\Big|_{E_7} = \frac{1}{\sqrt{2}} t_{ij}\Big|_{E_6} \ \ \to \ \ e_6 = \frac{e_7}{\sqrt{2}},
\end{equation}

\noindent
because

\begin{equation}
A_\mu\Big|_{E_7} = i e_7 A_\mu^{\bm{A}} t_{\bm{A}}\Big|_{E_7}\quad \to\quad A_\mu\Big|_{E_6} = ie_6 A_\mu^{\bm{A}} t_{\bm{A}}\Big|_{E_6}.
\end{equation}

For obtaining the coupling constant $e_1^{(6)}$ we write the branching rule for the representation $56(1)$ with respect to the subgroup $E_6\times U_1\times U_1$,

\begin{equation}
56(1)\Big|_{E_7\times U_1} = 27(1,1) + \,\xbar{27}\,(1,-1) + 1(1,3) + 1(1,-3)\Big|_{E_6\times U_1\times U_1},
\end{equation}

\noindent
and choose the charge with respect to the little group in the form

\begin{equation}\label{Little_Group_Charge}
Q_1^{(6)} = \frac{1}{2} \Big(- 3 Q_1^{(7)} + Q_2^{(7)} \Big).
\end{equation}

\noindent
From Eq. (\ref{E17_Coupling}) we see that the charge $Q_1^{(7)}$ is an eigenvalue of the operator $4\sqrt{3}\,t_1^{(7)}$, where $t_1^{(7)}$ is the generator of the $U_1$ factor in $E_7\times U_1$ which is normalized in the same way as the generators of the group $E_7$.

Let $t\Big|_{U_1\subset E_7}$ be the generator of the $U_1$ factor in the subgroup $E_6\times U_1 \subset E_7$ normalized in the same way as all $E_7$ generators. Then

\begin{equation}\label{U1_Of_E7}
t\Big|_{U_1\subset E_7} =
\frac{1}{12} \left(
\begin{array}{cccc}
3 & 0 & 0 & 0\\
0 & -3 & 0 & 0\\
0 & 0 & 1 & 0\\
0 & 0 & 0 & -1
\end{array}
\right)\quad
\text{acting on} \quad
\left(
\begin{array}{c}
1\\
1\\
27\\
\xbar{27}
\end{array}
\right),
\end{equation}

\noindent
because

\begin{equation}
\mbox{tr}\left(\Big(t\Big|_{U_1\subset E_7}\Big)^2\right) = \frac{1}{144} \Big(1\cdot 3^2 + 1\cdot (-3)^2 + 27\cdot 1^2 + 27\cdot 1^2\Big) = \frac{1}{2}.
\end{equation}

\noindent
Comparing Eq. (\ref{U1_Of_E7}) with the first branching rule in (\ref{E7_Branching_Rules}) we see that the charge $Q_2^{(7)}$ is an eigenvalue of the operator $12\, t\Big|_{U_1\subset E_7}$. Therefore, the little group charge (\ref{Little_Group_Charge}) corresponds to the operator

\begin{equation}
\frac{1}{2} \Big(- 3\cdot 4\sqrt{3}\, t_1^{(7)} + 12\, t\Big|_{U_1\subset E_7} \Big) = 12 \bigg(- \frac{\sqrt{3}}{2} t_1^{(7)} + \frac{1}{2} t\Big|_{U_1\subset E_7}\bigg).
\end{equation}

\noindent
In the right hand side the operator in the brackets is normalized in the same way as the generators of the group $E_7$. Therefore, the coefficient $12$ is equal to the ratio of the couplings $e_7$ and $e_1^{(6)}$, so that

\begin{equation}
e_1^{(6)} = \frac{e_7}{12}.
\end{equation}

Next, we construct the branching rule of $248$ with respect to the subgroup $E_6\times \underbrace{U_1}_{\beta_1^{(7)}}\times \underbrace{U_1}_{\beta_2^{(7)}}\subset E_7\times \underbrace{U_1}_{\beta_1^{(7)}}$,

\begin{eqnarray}
&&\hspace*{-7mm} 248\Big|_{E_8} = \Big[1(0,0) + 1(2,0) + 1(-2,0)\Big] + \Big[1(0,0) + 27(0,-2) + \,\xbar{27}\,(0,2) \nonumber\\
&&\hspace*{-7mm} + 78(0,0)\Big] + \Big[27(1,1) + \,\xbar{27}\,(1,-1) + 1(1,3) + 1(1,-3)\Big] + \Big[27(-1,1) \nonumber\\
&&\hspace*{-7mm} + \,\xbar{27}\,(-1,-1) + 1(-1,3) + 1(-1,-3)\Big]\bigg|_{E_6\times U_1\times U_1}
\end{eqnarray}

\noindent
and calculate the charge with respect to the little group for each term. As the result we obtain the decomposition

\begin{eqnarray}\label{248_E6}
&&\hspace*{-7mm} 248\Big|_{E_8} = 4\times 1(0) + 2\times 1(3) + 2\times 1(-3) + 2\times 27(-1) + 2\times\,\xbar{27}\,(1)  \vphantom{\Big|_{E_6}}\nonumber\\
&&\hspace*{-7mm} + 27(2) + \,\xbar{27}\,(-2) + 78(0)\Big|_{E_6\times U_1}.
\end{eqnarray}

\subsection{The representations for the further symmetry breaking}
\hspace*{\parindent}

The further investigation of the symmetry breaking is made similarly. Vacuum expectation values are acquired by the representations which are present in the branching rules of $248$ and contain singlets with respect to the non-Abelian components of the little group with nontrivial $U_1$ charges, namely, for $E_6\times U_1 \to SO_{10}\times U_1$

\begin{equation}
27\Big|_{E_6} = 1(4) + 10(-2)+16(1)\Big|_{SO_{10}\times U_1},
\end{equation}

\noindent
for $SO_{10}\times U_1 \to SU_5\times U_1$

\begin{equation}
16\Big|_{SO_{10}} = 1(-5)+\,\xbar{5}\,(3) + 10(-1)\Big|_{SU_5\times U_1},
\end{equation}

\noindent
and for $SU_5\times U_1 \to SU_3\times SU_2\times U_1$

\begin{equation}
10\Big|_{SU_5} = [1,1](6) + [\,\xbar{3}\,,1](-4) + [3,2](1)\Big|_{SU_3\times SU_2\times U_1}.
\end{equation}

\noindent
However, the further symmetry breaking can be made in a different ways because the $U_1$ charges of these representations can be different. For example, according to Eq. (\ref{248_E6}), the symmetry breaking $E_6\times U_1 \to SO_{10} \times U_1$ can be made either by vev of $27(-1)$ or by vev of $27(2)$ (and/or the corresponding conjugated representations). The various options which can appear in the considered symmetry breaking pattern will be analysed in section \ref{Section_Options}.

\section{Remaining relations between the coupling constants of the non-Abelian groups}
\hspace*{\parindent}\label{Section_Non_Abelian_Relations}

The remaining relations between the coupling constants for the non-Abelian groups can also be obtained by comparing the commutation relations for the corresponding generators using the explicit form of the embeddings. For instance, the $SO_{10}$ generators $(t_{ij})_{kl} = i\left(\delta_{ik}\delta_{jl} - \delta_{il}\delta_{jk}\right)/2$ normalized with the metric

\begin{equation}
g_{\bm{AB}} \to \delta_{ik}\delta_{jl} - \delta_{il}\delta_{jk};\qquad g^{\bm{AB}} \to \frac{1}{4}\Big(\delta_{ik}\delta_{jl} - \delta_{il}\delta_{jk}\Big)
\end{equation}

\noindent
satisfy the commutation relations

\begin{equation}
[t_{ij},t_{kl}] = \frac{i}{2}\Big(\delta_{il} t_{jk} - \delta_{jl} t_{ik} - \delta_{ik} t_{jl} + \delta_{jk} t_{il}\Big).
\end{equation}

\noindent
Comparing this equation with the corresponding relation for $E_6$ we conclude that

\begin{equation}
t_{ij}\Big|_{E_6} = \frac{1}{\sqrt{3}} t_{ij}\Big|_{SO_{10}} \ \ \to \ \ e_{10} = \frac{e_6}{\sqrt{3}},
\end{equation}

\noindent
because in this case

\begin{equation}
A_\mu\Big|_{E_6} = i e_6 A_\mu^{\bm{A}} t_{\bm{A}}\Big|_{E_6} \to A_\mu\Big|_{SO_{10}} = ie_{10} A_\mu^{\bm{A}} t_{\bm{A}}\Big|_{SO_{10}} = \frac{i}{2} e_{10} \big(A_\mu\big)_{ij} t_{ij}\Big|_{SO_{10}}.
\end{equation}

For constructing the embedding $U_5\subset SO_{10}$ we consider a complex 5-component column $z = x+iy$ such that

\begin{equation}
z\equiv x+iy \to \Omega_5 z = (B+iC)(x+iy) = (Bx -Cy) + i(By + Cx),
\end{equation}

\noindent
where $B$ and $iC$ are the real and purely imaginary parts of the $5\times 5$ matrix $\Omega_5 \in U_5$. The condition $\Omega_5^+ \Omega_5=1$ leads to the constraints

\begin{equation}\label{BC_Constraints}
B^T B + C^T C = 1;\qquad B^T C = C^T B.
\end{equation}

\noindent
The above transformation of $z$ can equivalently be presented as the transformation of the real 10-component column

\begin{equation}
\left(
\begin{array}{c}
x\\y
\end{array}
\right) \to
\left(
\begin{array}{cc}
B & -C\\
C & B
\end{array}
\right)
\left(
\begin{array}{c}
x\\y
\end{array}
\right).
\end{equation}

\noindent
Due to the constraints (\ref{BC_Constraints}) the matrix belongs to the group $SO_{10}$. (Its determinant is equal to 1 because the $U_5$ group manifold is connected.) Therefore, it is possible to write the properly normalized generators of $SO_{10}$ corresponding to the subgroup $SU_5$ in the form

\begin{equation}\label{SO_10_Generators}
t_A\bigg|_{SU_5\subset SO_{10}} = \left\{
\begin{array}{l}
{\displaystyle \frac{1}{\sqrt{2}}\left(
\begin{array}{cc}
t_{A,5} & 0\\
0 & t_{A,5}
\end{array}
\right) = \frac{1}{\sqrt{2}}\, T(t_{A,5}),} \\
\qquad\qquad\qquad\qquad\qquad\qquad \text{if $t_{A,5}$ is purely imaginary};\\
\vphantom{1}\\
{\displaystyle \frac{i}{\sqrt{2}}\left(
\begin{array}{cc}
0 & t_{A,5}\\
- t_{A,5} & 0
\end{array}
\right)= \frac{1}{\sqrt{2}}\, T(t_{A,5}),}\\
\qquad\qquad\qquad\qquad\qquad\qquad\qquad\qquad\quad\ \text{if $t_{A,5}$ is real},
\end{array}
\right.
\end{equation}

\noindent
where the generators of the $SU_5$ fundamental representation $t_{A,5}$ (with $A=1,\ldots,24$) are normalized by the condition

\begin{equation}
\mbox{tr}\big(t_{A,5} t_{B,5}\big) = \frac{1}{2}\delta_{AB}.
\end{equation}

\noindent
The properly normalized generator of the $U_1$ subgroup of $SO_{10}$ in this case takes the form

\begin{equation}
t\Big|_{U_1\subset SO_{10}} = -\frac{i}{\sqrt{20}} \left(
\begin{array}{cc}
0 & -1_5\\
1_5 & 0
\end{array}
\right).
\end{equation}

Due to the factor $1/\sqrt{2}$ in Eq. (\ref{SO_10_Generators}) the coupling constants for the groups $SO_{10}$ and $SU_5$ are related by the equation

\begin{equation}
e_5 = \frac{e_{10}}{\sqrt{2}}.
\end{equation}

Similarly, the last embedding $SU_3\times SU_2\times U_1\subset SU_5$

\begin{equation}
\omega_5 = \left(
\begin{array}{cc}
e^{-2i\beta_2^{(5)}} \omega_3 & 0 \\
0 &  e^{3i\beta_2^{(5)}} \omega_2
\end{array}
\right)
\end{equation}

\noindent
gives the well-known equation $e_2 = e_3= e_5$.

Thus, for the coupling constants corresponding to the non-Abelian groups we obtain the relations

\begin{equation}
e_2 = e_3 = e_5 = \frac{e_{10}}{\sqrt{2}} = \frac{e_6}{\sqrt{6}} = \frac{e_7}{\sqrt{12}} = \frac{e_8}{\sqrt{60}}.
\end{equation}

\section{The coupling constants for the $U_1$ groups and options for the further symmetry breaking}
\hspace*{\parindent}\label{Section_Options}

Next, we need to calculate all coupling constants corresponding to all $U_1$ groups present in the symmetry breaking pattern. For the symmetry breaking $G\times U_1\to H\times U_1$ this can be done according to the following algorithm:

1. First, it is necessary to construct the decomposition of the representation which acquires vev with respect to the subgroup $H\times U_1\times U_1 \subset G\times U_1$.

2. Next, one should find the expression for the little group charge. At all steps except for the last one it is chosen in such a way that this charge takes minimal possible integer values. At the last step the charge normalization is chosen so that the maximal number of MSSM representations has correct $U_1$ hypercharges.

3. After that, we construct the generators of the group $U_1\times U_1$ normalized in the same way as the generators of the group $G$.

4. Next, we construct the generator of the little group and extract from it the operator normalized in the same way as the generators of the group $G$. The coefficient before it gives the ratio $e_G/e_1^{(H)}$.

With the help of this algorithm for each option of the symmetry breaking we obtain a sequence of the $U_1$ charges. For each of them finally we calculate $\mbox{tg}\,\theta_W = e_1^{(Y)}/e_2$.

\begin{table}[h]
\hspace*{1mm}
\begin{tabular}{|c|c|c|c|c|}
\hline
Option & $E_6\times U_1$ & $SO_{10}\times U_1$ & $SU_5\times U_1\vphantom{\Big(}$ & $\sin^2\theta_W$\\
\hline
\bf{B-1-1-1} & $27(-1)\Big|_{E_6\times U_1}$ & $16(-1)\Big|_{SO_{10}\times U_1}$ & $10(-1)\Big|_{SU_5\times U_1}\vphantom{\Bigg(}$ & $\bm{3/8}$\\
\hline
B-1-1-2 & $27(-1)\Big|_{E_6\times U_1}$ & $16(-1)\Big|_{SO_{10}\times U_1}$ & $10(4)\Big|_{SU_5\times U_1}\vphantom{\Bigg(}$ & $3/5$\\
\hline
B-1-2-1 & $27(-1)\Big|_{E_6\times U_1}$ & $16(3)\Big|_{SO_{10}\times U_1}$ & $10(-2)\Big|_{SU_5\times U_1}\vphantom{\Bigg(}$ & $3/5$ \\
\hline
B-1-2-2 & $27(-1)\Big|_{E_6\times U_1}$ & $16(3)\Big|_{SO_{10}\times U_1}$ & $10(3)\Big|_{SU_5\times U_1}\vphantom{\Bigg(}$ & $3/4$\\
\hline
B-2-1-1 & $27(2)\Big|_{E_6\times U_1}$ & $16(1)\Big|_{SO_{10}\times U_1}$ & $10(-2)\Big|_{SU_5\times U_1}\vphantom{\Bigg(}$ & $3/5$\\
\hline
B-2-1-2 & $27(2)\Big|_{E_6\times U_1}$ & $16(1)\Big|_{SO_{10}\times U_1}$ & $10(3)\Big|_{SU_5\times U_1}\vphantom{\Bigg(}$ & $3/4$\\
\hline
\end{tabular}
\caption{Various options for the symmetry breaking $E_6\times U_1 \to SO_{10}\times U_1\to SU_5\times U_1\to SU_3\times SU_2\times U_1$}\label{Table_Further_Breaking}
\end{table}

All options for the symmetry breaking and the corresponding values of $\sin^2\theta_W$ obtained according to the procedure described above are presented in Table \ref{Table_Further_Breaking}. The various options appear because the scalar fields in the representations $27$ of $E_6$, $16$ of $SO_{10}$, and $10$ of $SU_5$ can have different $U_1$ charges. However, among these options there is the only one (denoted by B-1-1-1) which gives the correct value of the Weinberg angle. Moreover, this is the only option that contains all representations needed for the accommodation of all chiral MSSM superfields, because in this case the branching rule for the representation $248$ of $E_8$ with respect to the MSSM gauge group $SU_3\times SU_2\times U_1$ is \cite{Stepanyantz:2023vat}

\begin{eqnarray}
&&\hspace*{-5mm} 248\Big|_{E_8} = 25\times\bm{[1,1](0)} + 5\times[1,1](1) +5\times\bm{[1,1](-1)} + [1,3](0) \nonumber\\
&&\hspace*{-5mm} + 10\times\bm{[1,2](1/2)} + 10\times\bm{[1,2](-1/2)} +10\times \bm{[3,1](-1/3)}  \vphantom{\Big(}\nonumber\\
&&\hspace*{-5mm} + 10\times[\,\xbar{3}\,,1](1/3) + 5\times \bm{[3,1](2/3)} + 5\times [\,\xbar{3}\,,1](-2/3)  +[3,2](-5/6)  \vphantom{\Big(}\nonumber\\
&&\hspace*{-5mm} + [\,\xbar{3}\,,2](5/6) + 5\times [3,2](1/6) + 5\times \bm{[\,\xbar{3}\,,2](-1/6)} + [8,1](0) \Big|_{SU_3\times SU_2\times U_1},\nonumber\\
\end{eqnarray}

\noindent
where the representations needed for accommodating the MSSM superfields are maked by the bold font. Also it is interesting to note that B-1-1-1 corresponds to the minimal possible absolute values of the $U_1$ charges of the above mentioned representations $27$, $16$, and $10$. The values of coupling constants for all steps of symmetry breaking for the option B-1-1-1 are presented in Table \ref{Table_B111}.

\begin{table}[h]
\hspace*{3mm}
\begin{tabular}{|c|c|c|c|}
\hline
Group $\vphantom{\bigg(}$ & Vev & $e_G$ & $e^{(G)}_1$ \\
\hline
$E_8$ $\vphantom{\bigg(}$ & 248 & $e_8$ & $-$ \\
\hline
$E_7\times U_1$ $\vphantom{\bigg(}$ & $56(\pm 1)$ & $e_7=e_8/\sqrt{5}$ & $e_1^{(7)} = e_7/4\sqrt{3}$ \\
\hline
$E_6\times U_1$ $\vphantom{\bigg(}$ & $27(-1);\ \,\xbar{27}\,(1)$ & $e_6=e_7/\sqrt{2}$ & $e_1^{(6)}=e_6/6\sqrt{2}$ \\
\hline
$SO_{10}\times U_1$ $\vphantom{\bigg(}$ & $16(-1);\ \,\xbar{16}\,(1)$ & $e_{10} = e_6/\sqrt{3}$ & $e_1^{(10)} = e_{10}/4\sqrt{3}$\\
\hline
$SU_{5}\times U_1$ $\vphantom{\bigg(}$ & $10(-1);\ \,\xbar{10}\,(1)$ & $e_{5} = e_{10}/\sqrt{2}$ & $e_1^{(5)}=e_{5}/2\sqrt{10}$\\
\hline
$SU_3\times SU_2\times U_1$ $\vphantom{\bigg(}$ & $[1,2](\pm 1/2)$ & $e_3 = e_2 =e_5\vphantom{\Big(}$ & $e_1^{(Y)} = e_{5} \sqrt{3/5}$\\
\hline
\end{tabular}
\caption{Values of the coupling constants for the option B-1-1-1}\label{Table_B111}
\end{table}

The options B-1-1-2, B-1-2-1, and B-2-1-1 lead to the same value of the Weinberg angle $\sin^2\theta_W=3/5$ and to the same branching rule of the representation 248 with respect to $SU_3\times SU_2\times U_1$,

\begin{eqnarray}
&&\hspace*{-4mm} 248\Big|_{E_8} = 19\times\bm{[1,1](0)} + 8\times[1,1](1/2) +8\times[1,1](-1/2) + [1,3](0) \nonumber\\
&&\hspace*{-4mm} + 12\times [1,2](0) + 4\times\bm{[1,2](1/2)} + 4\times\bm{[1,2](-1/2)} +8\times [3,1](1/6)  \vphantom{\Big(}\nonumber\\
&&\hspace*{-4mm} + 8\times[\,\xbar{3}\,,1](-1/6) + 6\times \bm{[3,1](-1/3)} + 6\times [\,\xbar{3}\,,1](1/3) + \bm{[3,1](2/3)} \vphantom{\Big(}\nonumber\\
&&\hspace*{-4mm} + [\,\xbar{3}\,,1](-2/3) + 2\times [3,2](-1/3) + 2\times [\,\xbar{3}\,,2](1/3) +4\times [3,2](1/6)  \vphantom{\Big(}\nonumber\\
&&\hspace*{-4mm} + 4\times \bm{[\,\xbar{3}\,,2](-1/6)} + [8,1](0)\Big|_{SU_3\times SU_2\times U_1}.
\end{eqnarray}

\noindent
However, from this equation we see that the representation $[1,1](-1)$ needed for the superfields corresponding to the right charged leptons is absent in this case. Therefore, these options are not acceptable for phenomenology.

The options B-1-2-2 and B-2-1-2 also lead to the same value of the Weinberg angle $\sin^2\theta_W=3/4$ and to the same branching rule for the representation 248 with respect to $SU_3\times SU_2\times U_1$, which is written as

\begin{eqnarray}
&&\hspace*{-6mm} 248\Big|_{E_8} = 17\times\bm{[1,1](0)} + 9\times[1,1](1/3) + 9\times [1,1](-1/3) + [1,3](0)  \nonumber\\
&&\hspace*{-6mm} + \bm{[1,2](1/2)} + \bm{[1,2](-1/2)}  + 9\times [1,2](1/6) + 9\times [1,2](-1/6)  \vphantom{\Big(}\nonumber\\
&&\hspace*{-6mm} + 9\times [3,1](0) + 9\times[\,\xbar 3\,,1](0) + 3\times [3,1](1/3) + 3\times[\,\xbar{3}\,,1](-1/3) \vphantom{\Big(}\nonumber\\
&&\hspace*{-6mm} +3\times \bm{[3,1](-1/3)} + 3\times[\,\xbar{3}\,,1](1/3) + 3\times [3,2](-1/6) + 3\times [\,\xbar{3}\,,2](1/6) \vphantom{\Big(}\nonumber\\
&&\hspace*{-6mm} + 3\times [3,2](1/6) + 3\times \bm{[\,\xbar{3}\,,2](-1/6)} + [8,1](0)\Big|_{SU_3\times SU_2\times U_1}.
\end{eqnarray}

\noindent
In this case there are no representations $[1,1](-1)$ needed for the superfields corresponding to the right charged leptons and no representations $[3,1](2/3)$ corresponding to the right upper quarks. Therefore, all these options are also not acceptable for phenomenology.

\section{Conclusion}
\hspace*{\parindent}

Using the group theory we analyzed a possibility of the symmetry breaking pattern

\begin{equation}
E_8\to E_7\times U_1 \to E_6\times U_1 \to SO_{10}\times U_1 \to SU_5 \times U_1 \to SU_3 \times SU_2 \times U_1
\end{equation}

\noindent
provided that only parts of the representation $248$ can acquire vacuum expectation values. Also we assume that all $U_1$ groups in the considered chain are different. We have found that in this case there are 6 different options for the symmetry breaking, and the only one of them leads to the correct value of the Weinberg angle and produces all representations needed for the MSSM chiral matter superfields. It is interesting that this option corresponds to the minimal absolute values of all $U_1$ charges of the fields responsible for the symmetry breaking.

The representation $248$ of the group $E_8$ is fundamental and adjoint simultaneously, and, evidently, more than one $248$ representations are needed for realizing the symmetry breaking considered in this paper. Therefore, there is an interesting possibility \cite{Thomas:1985be} to use for the Grand Unification a finite theory obtained from ${\cal N}=4$ SYM with the group $E_8$ by adding some terms which break extended supersymmetry and do not break the finiteness (proved in \cite{Sohnius:1981sn,Grisaru:1982zh,Howe:1983sr,Mandelstam:1982cb,Brink:1982pd}). The existence of such terms was demonstrated in \cite{Parkes:1982tg,Parkes:1983ib,Parkes:1983nv}. However, here we did not study dynamics of the considered symmetry breaking pattern and made the investigation only using the group theory methods. This can be the prospect of future research. Making it, one should also take into account some other similar symmetry breaking patterns like

\begin{equation}
E_8\to E_7\times U_1 \to E_6\times U_1 \to SO_{10} \to SU_5 \times U_1 \to SU_3 \times SU_2 \times U_1,
\end{equation}

\noindent
etc. Although at present it is not clear if one can construct a phenomenologically acceptable theory based on the $E_8$ group, this possibility is worth considering.

%% The bibliography section


\begin{thebibliography}{99}

%\cite{Mohapatra:1986uf}
\bibitem{Mohapatra:1986uf}
R.N. Mohapatra: \emph{Unification and Supersymmetry. The Frontiers of Quark - Lepton Physics: The Frontiers of Quark-Lepton Physics}, Springer, 2002.
%doi:10.1007/978-1-4757-1928-4

%\cite{Minahan:1989vd}
\bibitem{Minahan:1989vd}
J.A. Minahan, P. Ramond, R.C. Warner: A Comment on Anomaly Cancellation in the Standard Model, Phys. Rev. D \textbf{41} (1990), 715--716.
%doi:10.1103/PhysRevD.41.715

%\cite{Bilal:2008qx}
\bibitem{Bilal:2008qx}
A. Bilal: Lectures on Anomalies: arXiv:0802.0634 [hep-th].

%\cite{Georgi:1974sy}
\bibitem{Georgi:1974sy}
H. Georgi, S.L.Glashow: Unity of All Elementary Particle Forces, Phys. Rev. Lett. \textbf{32} (1974), 438.
%doi:10.1103/PhysRevLett.32.438

%\cite{Dimopoulos:1981zb}
\bibitem{Dimopoulos:1981zb}
S. Dimopoulos, H.Georgi: Softly Broken Supersymmetry and SU(5), Nucl. Phys. B \textbf{193} (1981), 150--162.
%doi:10.1016/0550-3213(81)90522-8

%\cite{Sakai:1981gr}
\bibitem{Sakai:1981gr}
N. Sakai: Naturalness in Supersymmetric Guts, Z. Phys. C \textbf{11} (1981), 153--157.
%doi:10.1007/BF01573998

%\cite{Ellis:1990wk}
\bibitem{Ellis:1990wk}
J.R. Ellis, S. Kelley, D.V. Nanopoulos: Probing the desert using gauge coupling unification, Phys. Lett. B \textbf{260} (1991), 131--137.
%doi:10.1016/0370-2693(91)90980-5

%\cite{Amaldi:1991cn}
\bibitem{Amaldi:1991cn}
U. Amaldi, W.de Boer, H. Furstenau: Comparison of grand unified theories with electroweak and strong coupling constants measured at LEP, Phys. Lett. B \textbf{260} (1991), 447-455.
%doi:10.1016/0370-2693(91)91641-8

%\cite{Langacker:1991an}
\bibitem{Langacker:1991an}
P. Langacker, M.X. Luo: Implications of precision electroweak experiments for $M_t$, $\rho_{0}$, $\sin^2\theta_W$ and grand unification, Phys. Rev. D \textbf{44} (1991), 817--822.
%doi:10.1103/PhysRevD.44.817

%\cite{Fritzsch:1974nn}
\bibitem{Fritzsch:1974nn}
H. Fritzsch, P.Minkowski: Unified Interactions of Leptons and Hadrons, Annals Phys. \textbf{93} (1975), 193-266.
%doi:10.1016/0003-4916(75)90211-0

%\cite{Georgi:1974my}
\bibitem{Georgi:1974my}
H. Georgi: The State of the Art\textemdash{}Gauge Theories, AIP Conf. Proc. \textbf{23} (1975), 575--582.
%doi:10.1063/1.2947450

%\cite{Workman:2022ynf}
\bibitem{Workman:2022ynf}
R.L. Workman \textit{et al.} [Particle Data Group]: Review of Particle Physics, PTEP \textbf{2022} (2022), 083C01.
%doi:10.1093/ptep/ptac097

%\cite{Barr:1981qv}
\bibitem{Barr:1981qv}
S.M. Barr: A New Symmetry Breaking Pattern for SO(10) and Proton Decay, Phys. Lett. B \textbf{112} (1982), 219--222.
%doi:10.1016/0370-2693(82)90966-2

%\cite{Antoniadis:1987dx}
\bibitem{Antoniadis:1987dx}
I. Antoniadis, J.R. Ellis, J.S. Hagelin, D.V. Nanopoulos: Supersymmetric Flipped SU(5) Revitalized, Phys. Lett. B \textbf{194} (1987), 231--235.
%doi:10.1016/0370-2693(87)90533-8

%\cite{Campbell:1987eb}
\bibitem{Campbell:1987eb}
B.A. Campbell, J.R. Ellis, J.S. Hagelin, D.V. Nanopoulos: K.A.Olive, Supercosmology revitalized, Phys. Lett. B \textbf{197} (1987), 355--362.
%doi:10.1016/0370-2693(87)90400-X

%\cite{Ellis:1988tx}
\bibitem{Ellis:1988tx}
J.R. Ellis, J.S. Hagelin, S. Kelley, D.V. Nanopoulos: Aspects of the Flipped Unification of Strong, Weak and Electromagnetic Interactions, Nucl. Phys. B \textbf{311} (1988), 1--34.
%doi:10.1016/0550-3213(88)90141-1

%\cite{Masiero:1982fe}
\bibitem{Masiero:1982fe}
A. Masiero, D.V. Nanopoulos, K. Tamvakis, T.Yanagida: Naturally Massless Higgs Doublets in Supersymmetric SU(5), Phys. Lett. B \textbf{115} (1982), 380--384.
%doi:10.1016/0370-2693(82)90522-6

%\cite{Grinstein:1982um}
\bibitem{Grinstein:1982um}
B. Grinstein: A Supersymmetric SU(5) Gauge Theory with No Gauge Hierarchy Problem, Nucl. Phys. B \textbf{206} (1982), 387--396.
%doi:10.1016/0550-3213(82)90275-9

%\cite{Hisano:1994fn}
\bibitem{Hisano:1994fn}
J. Hisano, T. Moroi, K. Tobe, T.Yanagida: Suppression of proton decay in the missing partner model for supersymmetric SU(5) GUT, Phys. Lett. B \textbf{342} (1995), 138--144.
%doi:10.1016/0370-2693(94)01342-A
%[arXiv:hep-ph/9406417 [hep-ph]].

%\cite{Ellis:2020qad}
\bibitem{Ellis:2020qad}
J. Ellis, M.A.G. Garcia, N. Nagata, D.V. Nanopoulos, K.A. Olive: Proton Decay: Flipped vs Unflipped SU(5), JHEP \textbf{05} (2020), 021.
%doi:10.1007/JHEP05(2020)021
%[arXiv:2003.03285 [hep-ph]].

%\cite{Mehmood:2020irm}
\bibitem{Mehmood:2020irm}
M. Mehmood, M.U. Rehman, Q.Shafi: Observable proton decay in flipped SU(5), JHEP \textbf{02} (2021), 181.
%doi:10.1007/JHEP02(2021)181
%[arXiv:2010.01665 [hep-ph]].

%\cite{Haba:2021rzs}
\bibitem{Haba:2021rzs}
N. Haba, T. Yamada: Moderately suppressed dimension-five proton decay in a flipped SU(5) model, JHEP \textbf{01} (2022), 061.
%doi:10.1007/JHEP01(2022)061
%[arXiv:2110.01198 [hep-ph]].

%\cite{Ellis:2021vpp}
\bibitem{Ellis:2021vpp}
J. Ellis, J.L. Evans, N.Nagata, D.V.Nanopoulos, K.A. Olive: Flipped SU(5) GUT phenomenology: proton decay and $\mathbf {g_\mu - 2}$, Eur. Phys. J. C \textbf{81} (2021) 1109.
%doi:10.1140/epjc/s10052-021-09896-x
%[arXiv:2110.06833 [hep-ph]].

%\cite{Gursey:1975ki}
\bibitem{Gursey:1975ki}
F. Gursey, P. Ramond, P.Sikivie: A Universal Gauge Theory Model Based on E6, Phys. Lett. B \textbf{60} (1976), 177--180.
%doi:10.1016/0370-2693(76)90417-2

%\cite{King:2020ldn}
\bibitem{King:2020ldn}
S.F. King, S. Moretti, R.Nevzorov: A Review of the Exceptional Supersymmetric Standard Model, Symmetry \textbf{12} (2020) no.4, 557.
%doi:10.3390/sym12040557
%[arXiv:2002.02788 [hep-ph]].

%\cite{Konshtein:1980km}
\bibitem{Konshtein:1980km}
S.E. Konshtein, E.S. Fradkin: Asymptotically supersymmetric model of unified interaction based on E8 (in Russian), Pisma Zh. Eksp. Teor. Fiz. \textbf{32} (1980), 575--578.

%\cite{Baaklini:1980uq}
\bibitem{Baaklini:1980uq}
N.S. Baaklini: Supergrand unification in E8, Phys. Lett. B \textbf{91} (1980), 376--378.
%doi:10.1016/0370-2693(80)90999-5

%\cite{Baaklini:1980fv}
\bibitem{Baaklini:1980fv}
N.S. Baaklini: Supersymmetric exceptional gauge unification, Phys. Rev. D \textbf{22} (1980), 3118--3127.
%doi:10.1103/PhysRevD.22.3118

%\cite{Bars:1980mb}
\bibitem{Bars:1980mb}
I.Bars, M.Gunaydin: Grand Unification With the Exceptional Group E8, Phys. Rev. Lett. \textbf{45} (1980), 859--862.
%doi:10.1103/PhysRevLett.45.859

%\cite{Koca:1981xd}
\bibitem{Koca:1981xd}
M. Koca: On tumbling E8, Phys. Lett. B \textbf{107} (1981), 73--76.
%doi:10.1016/0370-2693(81)91150-3

%\cite{Thomas:1985be}
\bibitem{Thomas:1985be}
S. Thomas: Softly broken N=4 and E8, J. Phys. A \textbf{19} (1986), 1141--1149.
%doi:10.1088/0305-4470/19/7/016

%\cite{Slansky:1981yr}
\bibitem{Slansky:1981yr}
R. Slansky: Group Theory for Unified Model Building, Phys. Rept. \textbf{79} (1981), 1--128.
%doi:10.1016/0370-1573(81)90092-2

%\cite{Stepanyantz:2023vat}
\bibitem{Stepanyantz:2023vat}
K. Stepanyantz: The gauge coupling unification in the flipped $E_8$ GUT, arXiv:2305.01295 [hep-ph].

%\cite{Green:1987sp}
\bibitem{Green:1987sp}
M.B. Green, J.H. Schwarz, E. Witten: \emph{Superstring Theory. Vol. 1: Introduction}, Cambridge University Press, Cambridge, 1988.
%ISBN 978-0-521-35752-4

%\cite{Sohnius:1981sn}
\bibitem{Sohnius:1981sn}
M.F. Sohnius, P.C. West: Conformal Invariance in N=4 Supersymmetric Yang-Mills Theory, Phys. Lett. B \textbf{100} (1981), 245.
%doi:10.1016/0370-2693(81)90326-9

%\cite{Grisaru:1982zh}
\bibitem{Grisaru:1982zh}
M.T. Grisaru,  W.Siegel: Supergraphity. 2. Manifestly Covariant Rules and Higher Loop Finiteness, Nucl. Phys. B \textbf{201} (1982), 292
[erratum: Nucl. Phys. B \textbf{206} (1982), 496].
%doi:10.1016/0550-3213(82)90282-6

%\cite{Howe:1983sr}
\bibitem{Howe:1983sr}
P.S. Howe, K.S. Stelle, P.K. Townsend: Miraculous Ultraviolet Cancellations in Supersymmetry Made Manifest, Nucl. Phys. B \textbf{236} (1984), 125--166.
%doi:10.1016/0550-3213(84)90528-5

%\cite{Mandelstam:1982cb}
\bibitem{Mandelstam:1982cb}
S. Mandelstam: Light Cone Superspace and the Ultraviolet Finiteness of the N=4 Model, Nucl. Phys. B \textbf{213} (1983), 149--168.
%doi:10.1016/0550-3213(83)90179-7

%\cite{Brink:1982pd}
\bibitem{Brink:1982pd}
L. Brink, O. Lindgren, B.E.W. Nilsson: N=4 Yang-Mills Theory on the Light Cone, Nucl. Phys. B \textbf{212} (1983), 401-412.
%doi:10.1016/0550-3213(83)90678-8

%\cite{Parkes:1982tg}
\bibitem{Parkes:1982tg}
A.J. Parkes, P.C. West: $N=1$ Supersymmetric Mass Terms in the $N=4$ Supersymmetric {Yang-Mills} Theory, Phys. Lett. B \textbf{122} (1983), 365--367.
%doi:10.1016/0370-2693(83)91583-6

%\cite{Parkes:1983ib}
\bibitem{Parkes:1983ib}
A. Parkes, P.C. West: Finiteness and Explicit Supersymmetry Breaking of the $N=4$ Supersymmetric {Yang-Mills} Theory, Nucl. Phys. B \textbf{222} (1983), 269--284.
%doi:10.1016/0550-3213(83)90637-5

%\cite{Parkes:1983nv}
\bibitem{Parkes:1983nv}
A. Parkes, P.C. West: Explicit Supersymmetry Breaking Can Preserve Finiteness in Rigid $N=2$ Supersymmetric Theories, Phys. Lett. B \textbf{127} (1983), 353--359.
%doi:10.1016/0370-2693(83)91016-X

\end{thebibliography}
\end{document}